\documentstyle[12pt,epsf,epsfig]{article}
\newcommand{\lab}{\label}

\newcommand{\bc}{\begin{center}}
\newcommand{\ec}{\end{center}}
\newcommand{\be}{\begin{equation}}
\newcommand{\ee}{\end{equation}}
\newcommand{\bea}{\begin{eqnarray}}
\newcommand{\eea}{\end{eqnarray}}
\newcommand{\bs}{\begin{subequations}}
\newcommand{\es}{\end{subequations}}
\newcommand{\beq}{\begin{eqalignno}}
\newcommand{\eeq}{\end{eqalignno}}

\def\lab{\label}
\begin{document}

\begin{center}
{\Large\bf Linear and Nonlinear Regime of a Random Resistor Network
Under Biased Percolation } \\

\vspace{0.5cm}

C. Pennetta, E. Alfinito and L. Reggiani \\
\vspace{0.6cm}
{\it INFM National Nanotechnology Laboratory and Dipartimento di Ingegneria 
\\dell'Innovazione, Universit\`a di Lecce, Via Arnesano, 73100 Lecce, Italy}
\end{center}

\begin{abstract}
We investigate the steady state of a two-dimensional random resistor 
network subjected to two competing biased percolations as a function of
the bias strength. The properties of the linear and nonlinear regimes 
are studied by means of Monte Carlo simulations.
In constant current conditions, a scaling relation is found between
$<R>/<R>_0$ and $I/I_0$, where  $<R>$ is the average network resistance,
$<R>_0$ the Ohmic resistance and $I_0$ an appropriate
threshold value for the onset of nonlinearity. 
A similar scaling relation is found also
for the relative variance of resistance fluctuations.
These results are in good agreement with electrical breakdown
measurements performed in composite materials.

\end{abstract}

The study of electrical and mechanical stability of disordered systems
is attracting a considerable interest because of its implications on 
both the material technology\cite{bardhan,ohring} and fundamental aspects 
related to the understanding of the behavior of these systems 
\cite{bardhan,hermann,havlin,stanley}. Indeed, the application of a finite 
stress (electrical or mechanical) to a disordered material generally implies a 
nonlinear response, which ultimately leads to an irreversible breakdown in 
the high stress limit \cite{bardhan,hermann}.
Such catastrophic phenomena have been successfully studied by using
percolation theories \cite{bardhan,hermann,havlin,stanley,stauffer}. 
In particular, large attention has been devoted to the critical exponents 
describing the resistance and its relative noise in terms of the property of 
the medium \cite{bardhan,havlin,stauffer,rammal85}. However, few attempts 
have been made so far to describe the behavior of a disordered media over the 
full range of the applied stress \cite{sornette96,mukherjee}.
On the other hand, such a study can provide important information about 
precursor effects and, more generally, about the predictability of 
breakdown \cite{sornette96,mukherjee}. Therefore, a satisfactory understanding
of breakdown phenomena is demanding \cite{bardhan,hermann}.

Here, our aim is to present a percolative model of sufficient generality to 
address the above issues. Our goal is to provide a theoretical framework to 
study response and fluctuation phenomena under linear and nonlinear regimes 
in a wide class of disordered systems. To this purpose, we analyze the 
evolution of a random resistor network (RRN) in which two competing 
percolative processes are present, defect generation and defect recovery.
These processes, which determine the values of the elementary network 
resistances, are driven by the joint effect of an electrical bias and of the 
heat exchange with a thermal bath. The bias is set up by applying a constant 
current. Monte Carlo (MC) simulations are performed to investigate the network
evolution in the full range of bias values. Indeed, depending on the bias 
strength, an irreversible failure or a steady state of the RRN can be 
achieved. By focusing on the steady state, we analyze the behavior of the
average network resistance, $<R>$, and the properties of the resistance
fluctuations, as a function of the bias.

The paper is organized as follows. In Sect. 1 we briefly describe the model
used. Section 2 presents the results of the MC simulations for the resistance
and its fluctuations. The main conclusions are drawn in Sect. 3
\section{The model} 
We study a two-dimensional random resistor network of total resistance $R$,
made of $N_{tot}$ resistors, each  of resistance $r_n$, disposed on a square
lattice. We take a square geometry, $N \times N$, where $N$ determines the
linear size of the lattice. For the comparison with resistance measurements 
of thin films, the value of $N$ can be related to the ratio between the size 
of the sample and the average size of the grains composing the sample.
An external bias represented by a constant current $I$, is applied to the RRN 
through electrical contacts realized by perfectly conducting bars at the left 
and right hand sides of the network. A current $i_n$ is then flowing through 
each resistor. The RRN interacts with a thermal bath at temperature $T_0$ and 
the resistances $r_n$ depend linearly on the local temperatures, $T_n$, as:
\begin{equation}
r_{n}(T_{n})=r_{0}[ 1 + \alpha (T_{n} - T_0)]
\label{eq:tcr}
\end{equation}
In this expression $\alpha$ is the temperature coefficient of the 
resistance, and $T_n$ is calculated by adopting a biased percolation 
model \cite{pen_prl_fail} as:
\begin{equation}
T_{n}=T_{0} + A \Bigl[ r_{n} i_{n}^{2} + {B \over N_{neig}}
\sum_{l=1}^{N_{neig}}  \Bigl( r_{l} i_{l}^2   - r_n i_n^2 \Bigr) \Bigr]
\label{eq:temp}
\end{equation}
Here, $N_{neig}$ is the number of first neighbours around the n{\em th}
resistor, the parameter $A$ describes the heat coupling of each resistor 
with the thermal bath and it determines the importance of Joule heating 
effects. The parameter $B$ is taken to be equal to $3/4$ to provide a uniform 
heating for the case of perfect network. In the initial state (corresponding 
to the perfect network with no heating) all the resistors are identical:
$r_n \equiv r_0$. Accordingly, the initial value of the resistance is
$R_0=r_0[N/(N+1)]$. Now, we assume that two competing percolative processes
act to determine the RRN evolution. The first process consists of generating
fully insulating defects (resistors with very high resistance, i.e. broken
resistors) with probability $W_{D,n}=exp[ -E_D/K_B T_n ]$, where $E_D$ is an
activation energy characteristic of the defect and $K_B$ the Boltzmann
constant \cite{pen_prl_fail}. The second  process consists of recovering the 
insulating defects with probability  $W_{R,n}=exp[ -E_R/K_B T_n ]$, where 
$E_R$ is an activation energy characteristic of this second process. 
For $A \neq 0$, Eq.~(\ref{eq:temp}) implies that both the processes 
(defect generation and defect recovery) are correlated percolations. Indeed, 
the probability of breaking (recovering) a resistor is higher in the so 
called ``hot spots'' of the RRN. On the other hand, for $A=0$ 
Eq.~(\ref{eq:temp}) yields $T_n \equiv T_0$, which corresponds to random 
percolations \cite{stauffer,pen_prl_stat}. The same is true for vanishing 
small bias values, when Joule heating effects are negligible. As a result of 
the competition between these two percolations the RRN reaches a steady state 
or fails completely with a critical fraction of defect $p_c$, corresponding 
to the percolation threshold \cite{stauffer}. In the first case, the
network resistance fluctuates around an average value $<R>$, while in the
second case, an irreversible breakdown occurs, i.e. $R$ diverges due to the
existence of at least one continuous path of defects between the upper and
lower sides of the network \cite{stauffer}.
We note that in the limit of a vanishing bias (random percolation) and
infinite lattices ($N \rightarrow \infty$), the expression:
$E_R < E_D + K_BT \ ln [1 + exp(-E_D/K_BT)]$
provides a sufficient condition for the existence of a steady state 
\cite{pen_prl_stat}, while it represents only a necessary condition in the 
case of biased percolation.

The evolution of the RRN is obtained by MC  simulations carried out according
to the following iterative procedure. (i) Starting from the perfect lattice
with given local currents, the local temperatures $T_n$ are calculated
according to Eq.~(\ref{eq:temp}); (ii) the defects are generated with
probability $W_{D,n}$ and  the resistances of the unbroken resistors are 
changed as specified by Eq.~(\ref{eq:tcr}); (iii) the currents $i_n$ are 
calculated by solving Kirchhoff's loop equations by the Gauss elimination 
method and the local temperatures are updated; (iv) the defects are recovered 
with probability $W_{R,n}$ and the temperature dependence of unbroken 
resistors is again accounted for; (v) $R$, $i_n$ and $T_n$ are finally 
calculated and the procedure is iterated from (ii) until one of the two 
following possibilities is achieved. In the first, electrical breakdown 
occurs (in numerical calculations we stop the iteration when $R$ increases 
over a factor of $10^3$ with respect to its initial value). In the second, 
the RRN reaches a steady state; in this case the iteration runs long enough to
allow a fluctuation analysis to be carried out. Each iteration step can be 
associated with an elementary time step on an appropriate time scale 
(to be calibrated with experiments).
In the simulations, as reasonable values of the parameters, we have taken:
$N =75$, $r_0=1 (\Omega)$, $\alpha = 10^{-3}$ (K$^{-1}$),
$A=5 \times 10^5$ (K/W) and $T_0=300 \ (K)$ if not stated otherwise.
Furthermore,  $E_{D} = 0.17\ (eV)$ and values of $E_{R}$ ranging 
from $0.026\  (eV)$ to $0.155 \ (eV)$ are used.  
The values of the external bias range from $10^{-4} \le I \le 2.8$ (A).
\section{Results}
\begin{figure}[t]
	\centerline{
	\epsfig{file=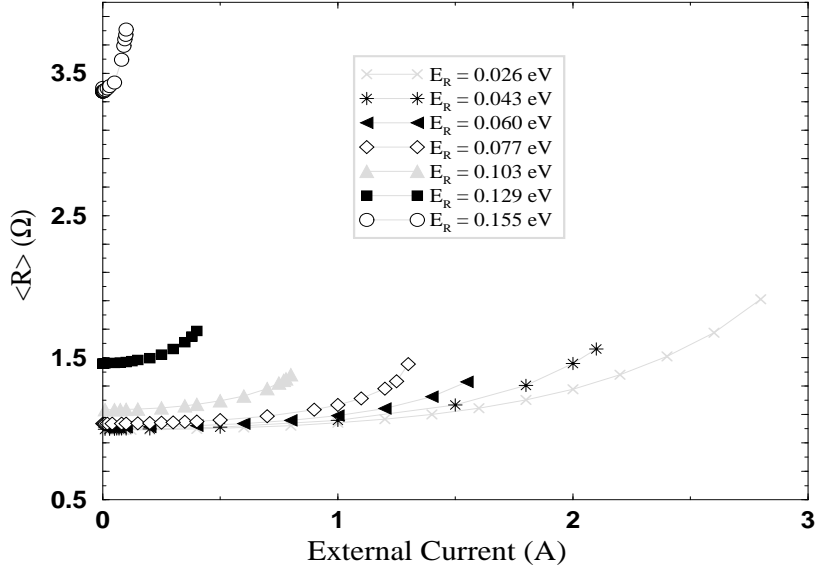,width=0.5
	\linewidth,height=0.7\linewidth,angle=-90}}
	\caption{\protect\small Average resistance versus 
the external current.  Each set of data
refers to the $E_{R}$ values shown in the legend.}
	\label{fig:fig1}
	\end{figure}
	\begin{figure}[t]
	\centerline{
	\epsfig{file=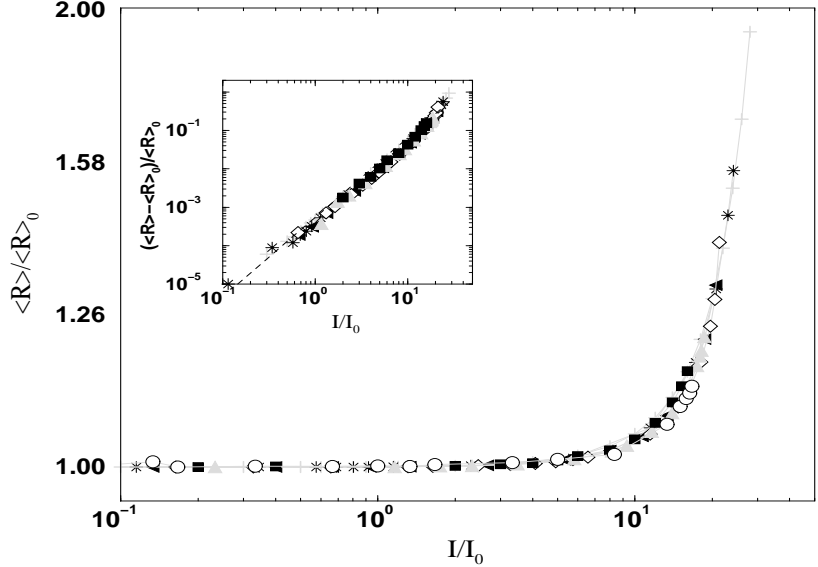,width=0.5
	\linewidth,height=0.7\linewidth,angle=-90}}
	\caption{\protect\small Log-log plot of the normalized 
	resistances versus the normalized bias. 
	Each set of data refers to different values of $E_{R}$ as reported
	in the legend of Fig. 1.
 The insert reports the quantity $\left(\langle R\rangle-\langle R_{0}
\rangle\right)/\langle R_{0}\rangle$ and the dotted line fits the
power law $y\propto x^{n}$ with $n=2$.} 
	\label{fig:fig2}
	\end{figure}
	\begin{figure}[t]
	\centerline{
	\epsfig{file=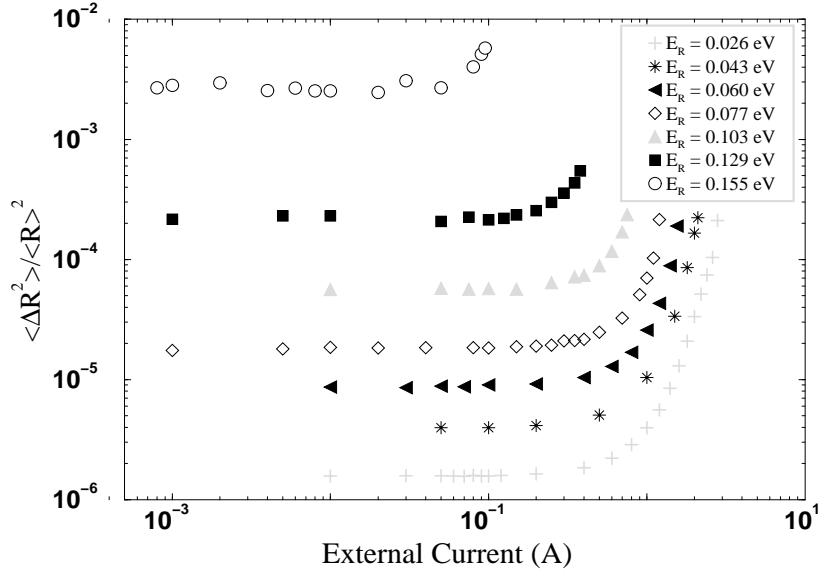,width=0.5
	\linewidth,height=0.7
	\linewidth,angle=-90}}
	\caption{\protect\small Log-log plot of the relative variance of 
	   resistance fluctuations versus the external current. 
           Each set of data refers to the $E_{R}$ values shown in the legend.}
	\label{fig:fig3}
	\end{figure}
\begin{figure}[t]
	\centerline{
	\epsfig{file=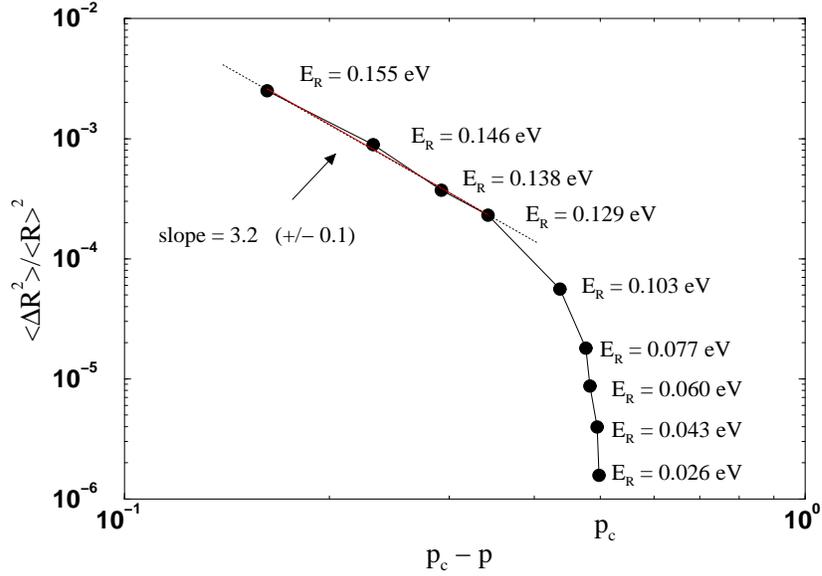,width=0.5
	\linewidth,height=0.7\linewidth,angle=-90}}
	\caption{\protect\small Log-log plot of the relative variance of 
	  resistance fluctuations versus $p_{c}-p$. All the values refer to 
          the linear regime.}
	\label{fig:fig4}
	\end{figure}
	\begin{figure}[t]
	\centerline{
	\epsfig{file=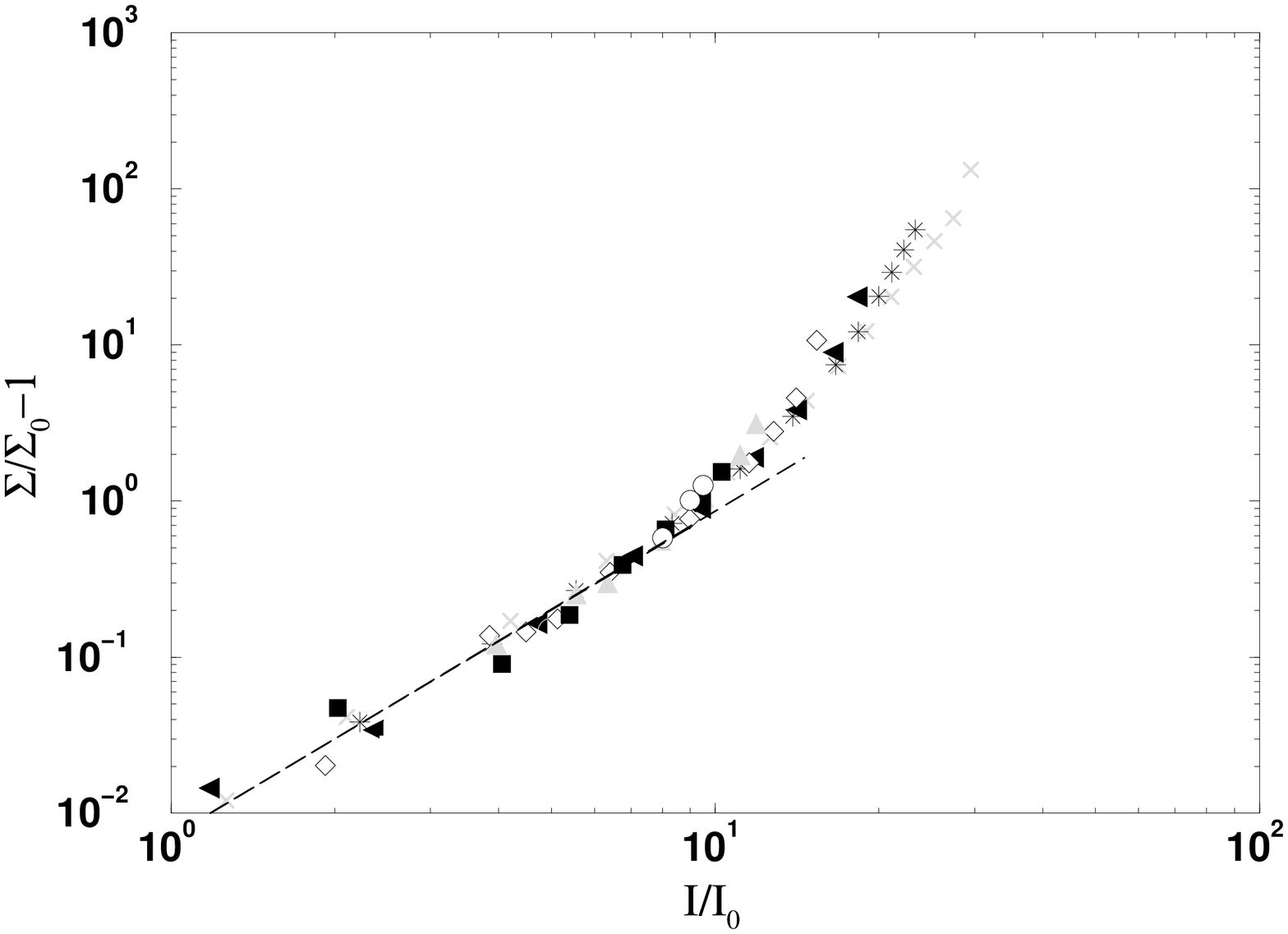,width=0.5
	\linewidth,height=0.7\linewidth,angle=-90}}
	\caption{\protect\small Log-log plot of $\Sigma/\Sigma_0-1$
          where $\Sigma/\Sigma_0$ is the relative variance of 
          resistance fluctuations normalized to the same quantity calculated 
          in the linear regime. The dashed line emphasizes the region of 
          validity of the quadratic power law.} 
	\label{fig:fig5}
	\end{figure}
Figure 1 reports the mean value of the RRN resistance as a function of
the external current for the values of $E_{R}$ reported in the figure. 
Each value of $<R>$ has been calculated by consider the time average on a 
single realization and then averaging over 20 realizations. The general 
behavior of each set of data shows a systematic increase of the resistance 
up to the breakdown of the network which occurs for currents greater than the 
breakdown value $I_b$ corresponding to the last point reported in the figure.
For each set of data, at increasing current values, we identify three main 
transport regimes, namely: a linear regime at the lowest currents, a nonlinear
regime at intermediate current values, and a pre-breakdown regime at the 
highest currents. The linear regime corresponds to a resistance which is 
independent of current (Ohmic response) and extends wider the lower the 
recovery energy. The nonlinear regime can be described by a quadratic 
dependence of the resistance in an appropriate range of current values 
starting from a threshold value $I_0$ and will be analyzed with details in 
the following. The pre-breakdown regime covers a narrow region just before 
the breakdown of the network which will be better analyzed on the basis of 
the resistance fluctuations. The fact that by increasing the recovery energy 
the breakdown occurs at lower currents is explained by the concomitant growth 
of the average number of defects, and thus of the disorder, inside the network
 which implies its lower stability against breakdown.

The similar feature of the data reported in Fig. 1 suggests the possibility
that all the curves belong to the same universality class. We define the value 
of the current $I_{0}$ by adopting the criterium of the departure of the
resistance from the linear response value of $0.05 \%$. Accordingly, Fig. 1 
has been redrawn on a normalized current scale $I/I_{0}$ as is reported in 
Fig. 2. This figure, by showing that all curves collapse into a single
one within an uncertainty of a few percent, clearly confirms the above 
suggestion. Furthermore, to obtain a better insight of the quadratic regime, 
in the insert we report the relative variation of resistance
$\left(\langle R\rangle-\langle R_{0}\rangle\right)/\langle R_{0}\rangle$ 
as a function of the normalized current. Here, we have found that the 
power law:  
\be
\frac{R(I)}{R_{0}} = g(\frac{I}{I_{0}}), \qquad {\rm with}\qquad
g(\frac{I}{I_{0}}) = 1+ \left(\frac{I}{I_{0}}\right)^{2},
\lab{MBH1}\ee
is well satisfied over about two decades of current values. The evidence of 
such a quadratic regime is in good agreement with recent experimental 
results \cite{mukherjee} concerning the electrical failure of carbon 
high-density polyethylene composites (HDPE). In particular, their failure 
transition is found to be of the first order which is a feature we have 
checked in the present model. From the study of HDPE samples with different
concentration of conductive component (carbon black) it was discovered that 
the ratio $R_{b}/R_{0}$ between the breakdown resistance $R_b$ and the Ohmic 
value $R_0$  is a constant independent of the value of the carbon fraction and 
the geometry of the sample, depending only on the nature of the conducting 
component. This result emerges, in our context, as a consequence of the
independence of the ratio $I/I_0$ on the initial resistance \cite{pen_physc}
and of Eq.(\ref{MBH1}). The electrical properties of these composites are 
thus nicely described by a RRN modelling. 

An important source of information concerning the physical properties of the 
stationary regime comes from the study of the resistance fluctuations. 
To this purpose we analyze in the following the variance of resistance
fluctuations as obtained by the simulations.
Figure 3 reports the relative variance of resistance fluctuations
$\langle \Delta R^{2} \rangle/\langle R \rangle^2 \equiv \Sigma$,  
with $\langle \Delta R^{2} \rangle = $ 
$\langle R^{2} \rangle - \langle R \rangle^{2}$,  as a function of the 
external current for the different values of recovery energy reported in the 
figure. The general behavior of different sets are similar to those reported
in Fig. 1 for the resistance. Also in this case we find a systematic increase 
of the variance at increasing current with the remarkable novelty that now 
the increase can be of several order of magnitude for the most ordered network.
By increasing the value of $E_{R}$, the curves start with higher values, in 
other words, the variance grows with $E_{R}$. Each curve is initially 
independent of current, which corresponds to the linear regime where the 
variance is an intrinsic property of the RRN. Then, the variance starts 
increasing with current according with the establishment of the non-linear 
regime driven by the external current and accounted for by the bias 
percolation. Finally, the breakdown of the network is achieved but in a much 
sharp way than for the case of the resistance. We note, that the results at 
the two highest values of the recovery energy (open circles and full squares 
in the figure) are noisy already in the linear regime due to an initial 
network which is significantly disordered and thus less stable than those 
corresponding to low values of $E_R$. 

The linear regime, occurring for small currents, when Joule heating effects 
are negligeable, practically corresponds to the competition of two random 
percolations \cite{stauffer}. We have thus investigated what kind of noise 
characterizes this random regime. Figure 4 reports the relative variance of 
resistance fluctuations $\Sigma$, as a function of the network defectiveness 
monitored by $p_{c} - p$ where $p_{c}=0.5$ is the critical value of the 
fraction of defects (i.e. the percolation threshold) for a square lattice and 
$p$ the average fraction of defects under stationary conditions.
The plotted data have been calculated for bias $I<I_{0}$ and each point refers
to the different value of $E_{R}$ reported in the figure. For recovery 
energies smaller than about $0.103 \ (eV)$ we obtain a quasi-perfect lattice, 
i.e. $p \ll p_{c}$, while for larger values, the fraction of defects become 
significant and $\Sigma$ grows following the scaling law: 
$\Sigma \propto (p_{c} - p)^{k}$. Our simulations give $k=3.2 \pm 0.1$, 
thus confirming previous findings obtained by taking the thermal coupling 
parameter $A=0$  \cite{pen_prl_stat}. We note, that in literature 
\cite{rammal85} the value $1.12 \pm 0.02$ has been found for the 
case of a flicker noise spectrum in disordered network subjected to random 
percolation. Therefore, the present noise should be interpreted as coming from 
the intrinsic source of fluctuations and with the property of
being  much more sensitive to the presence of disorder.

As for the case of the resistance, the similar features of the data reported 
in Fig. 3 suggest the possibility that all the curves belong to the same 
universality class. To this purpose, Fig. 5 reports on a log-log scale 
the quantity $\Sigma/\Sigma_{0}-1$ as a function of the normalized current 
$I/I_{0}$. Here $\Sigma_0$ indicates the variance of resistance fluctuations 
for the Ohmic regime. The data are the same of those in Fig. 3 and all the 
curves are found to collapse onto a single one. Looking in detail the data 
in Fig. 5, we find that beside the quadratic regime, higher order terms are 
present \cite{pen_physc}. Indeed, by analogy with the case of resistance, 
the quadratic regime is identified at the lowest current ($I < 10 \ I_{0}$) 
as emphasized by the continuous line. At high currents, ($I > 10 \ I_{0}$) 
a $4$-th order and eventually a $6$-th order regime can be identified. 
The presence of these high order terms can be accounted for by a unique
law which expands the variance of resistance fluctuations in an even power 
series of the external current. In particular, for large values of $E_{R}$ the 
breakdown happens for small $I_{b}$, so only the quadratic term
of the perturbative expansion, $\left(I/I_{0}\right)^{2}$, is appreciable.
By contrast, for small values of $E_{R}$ the breakdown happens for large
$I_{b}$  and terms of order higher than the second are detected.
We conclude that the noise is more sensitive than resistance to probe
the pre-breakdown region, since fluctuations are here dramatically amplified
with respect to negligible deviations of the resistance from the quadratic
regime.

\section{Conclusions}
We have studied the steady state of a two-dimensional random resistor 
network resulting from the simultaneous evolutions of two competing 
percolations biased from an external current. The two percolations consist
in a defect generation process and in a defect recovery one. 
The state of the network is investigated by Monte Carlo simulations over 
the full range of the applied stress. In particular, we have analyzed the
behavior of the average resistance and of the relative variance of resistance
fluctuations as a function of the external current and for different values
of the recovery activation energy. We have found that both these quantities
follow a scaling relation in terms of $I/I_0$, i.e. the ratio between the 
applied current and the current value corresponding to the onset of 
nonlinearity. The scaling exponents are found to be independent of the 
activation energy value, implying that all the failure processes belong to 
the same universality class. In the linear regime, the relative variance of 
resistance fluctuations has been analysed with respect to the average 
fraction of defects in the network and, at increasing disorder, a scaling law 
as a function of $p_c-p$ has been obtained. In the nonlinear regime, 
the relative variance of resistance fluctuations exhibits a quadratic 
dependence on the current while the presence of higher order terms 
characterizes the pre-breakdown region. The analysis of these higher order 
terms is left as a subject of further investigation. 

\vskip2pc\noindent 
Acknowledgements
\vskip1pc\noindent 
This research is performed within the STATE project of INFM.
Partial support is also provided by the MADNESS II project of the Italian
National Research Council and the ASI project N. 253.


\end{document}